\documentclass[aps,jcp,preprint,superscriptaddress]{revtex4}
\usepackage{graphicx}
\usepackage{amsmath}
\usepackage{dcolumn}
\usepackage{bm}

\begin{document}
\title{Cold and ultracold NH--NH collisions in magnetic fields}

\author{Liesbeth M.~C.~Janssen}
\affiliation{Radboud University Nijmegen, Institute for Molecules and Materials,
Heyendaalseweg 135, 6525 AJ Nijmegen, The Netherlands}
\author{Piotr S.~\.{Z}uchowski}
\affiliation{Department of Chemistry, Durham University, South Road, Durham, DH1 3LE, 
United Kingdom}
\author{Ad van der Avoird}
\affiliation{Radboud University Nijmegen, Institute for Molecules and Materials,
Heyendaalseweg 135, 6525 AJ Nijmegen, The Netherlands}
\author{Gerrit C.~Groenenboom}
\email[Electronic mail: ]{Gerritg@theochem.ru.nl}
\affiliation{Radboud University Nijmegen, Institute for Molecules and Materials,
Heyendaalseweg 135, 6525 AJ Nijmegen, The Netherlands}
\author{Jeremy M.~Hutson}
\email[Electronic mail: ]{J.M.Hutson@durham.ac.uk}
\affiliation{Department of Chemistry, Durham University, South Road, Durham, DH1 3LE, 
United Kingdom}

\date{\today}

\begin{abstract}
Elastic and spin-changing inelastic collision cross sections are presented for cold and
ultracold magnetically trapped NH. The cross sections are obtained from 
coupled-channel scattering calculations as a function of energy and magnetic field. We
specifically investigate the influence of the intramolecular spin-spin,
spin-rotation, and intermolecular magnetic dipole coupling on the collision
dynamics. It is shown that $^{15}$NH is a very suitable candidate for
evaporative cooling experiments.  The dominant trap-loss mechanism in the
ultracold regime originates from the intermolecular dipolar coupling term. At
higher energies and fields, intramolecular spin-spin coupling becomes
increasingly important. Our qualitative results and
conclusions are fairly independent of the exact form of the potential and
of the size of the channel basis set. 
\end{abstract}

\maketitle

\section{Introduction}
Cold ($T < 1$ K) and ultracold ($T < 1$ mK) molecules have attracted great
interest in recent years due to their potential applications in condensed
matter physics \cite{micheli:06}, quantum computing \cite{demille:02,andre:06},
high-precision spectroscopy \cite{lev:06,bethlem:09,tarbutt:09}, and physical
chemistry \cite{meerakker:05, gilijamse:06,gilijamse:07,campbell:08,krems:08}.
Experimental methods for producing ultracold molecular samples
include photoassociation \cite{jones:06} and Feshbach association
\cite{kohler:06}, in which molecules are formed by linking two ultracold atoms.
These indirect cooling techniques are, however, limited to molecules consisting
of two (alkali metal) atoms that can be efficiently pre-cooled to ultralow
temperatures. Direct cooling methods such as buffer-gas cooling
\cite{weinstein:98} and Stark deceleration \cite{bethlem:03}, in which
molecules are cooled from room temperature, are applicable
to a much wider range of species.

The NH radical is an excellent candidate for direct cooling experiments, due to
its relatively large magnetic and electric dipole moments. Ground-state
NH($X\,^3\Sigma^-$) has been successfully cooled with a helium buffer gas and
confined in a magnetic trap \cite{krems:03a,campbell:07,hummon:08,campbell:08}.
Metastable NH($a\,^1\Delta$) has been Stark decelerated and trapped in an
electrostatic field \cite{hoekstra:07}. Furthermore, an optical pumping scheme
has been developed to transfer the Stark-decelerated molecules to the
electronic ground state and subsequently accumulate them in a superimposed
magnetic trap \cite{meerakker:01}.

At present, direct cooling experiments for NH have achieved temperatures of a
few hundred mK. One of the key challenges is to produce
molecules in the ultracold regime, which requires a second-stage cooling
method. Examples of such methods include sympathetic cooling with an ultracold
co-trapped species \cite{soldan:09,wallis:09,wallis:10,zuchowski:10,hummon:10}
and evaporative cooling. The latter approach relies on elastic NH + NH
collisions that thermalize the gas as the magnetic trap depth is slowly
decreased. Inelastic collisions, however, can induce Zeeman relaxation and
consequently lead to trap loss. Therefore, in order for evaporative cooling to
work, elastic collisions should be much more efficient than inelastic ones,
typically by a few orders of magnitude \cite{ketterle:96}.
Evaporative cooling has already been achieved for buffer-gas
cooled metastable helium \cite{doret:09}, indicating that the densities
in buffer-gas cooling experiments are high enough to obtain a sufficiently
high elastic collision rate. This also provides hope for other buffer-gas
cooled species such as NH.

A collision complex consisting of two magnetically trapped NH molecules is in
the \mbox{$| S=2, M_S=2 \rangle$} state, with $S$ denoting the total
spin and $M_S$ its laboratory-frame projection.  Inelastic NH + NH collisions
can change the $M_S$ quantum number and/or the total spin of the dimer.  If the
total spin is flipped, the complex undergoes a transition from the quintet
state to the chemically reactive singlet or triplet states
\cite{dhont:05,janssen:09,lai:03}. Although unfavorable for evaporative
cooling, the latter two spin states may be relevant in the context of cold
controlled chemistry \cite{krems:08}.

Previous theoretical work by Kajita \cite{kajita:06} and Janssen \textit{et
al.}\ \cite{janssen:10} has shown that NH is a promising candidate for
molecular evaporative cooling, particularly the bosonic $^{15}$NH isotope.
Fermionic $^{14}$NH may also be cooled into the $\mu$K regime either by
applying an electric field \cite{kajita:06} or by preparing the molecules in
different hyperfine states \cite{janssen:10}.  It must be noted, however, that
the results of Kajita were obtained from approximate analytical methods in
which only the electric dipole-dipole and magnetic dipole-dipole interactions
were included.  The work of Ref.\ \cite{janssen:10} involved rigorous quantum
scattering calculations on a full 4-dimensional potential-energy surface,
but focused only on NH + NH collisions in zero field.

In this paper, we consider collisions between $^{15}$NH molecules in the
presence of an external magnetic field.  We present full quantum scattering
calculations on an accurate \textit{ab initio} quintet potential to investigate
the collision dynamics at low and ultralow temperatures. Intramolecular
spin-spin coupling, spin-rotation, and intermolecular magnetic dipolar coupling
are explicitly included in the calculations.  We will identify the dominant
trap-loss mechanisms and provide a detailed discussion of the dynamics.

\section{Calculations}
\label{sec:calc}
\subsection{Theory}
\label{subsec:theory}
We consider elastic and inelastic collisions between two identical
NH($X\,^3\Sigma^-$) molecules in the presence of a magnetic field.  Since
bosonic $^{15}$NH is more suitable for evaporative cooling that fermionic
$^{14}$NH, we will focus only on the $^{15}$NH isotopologue.  We treat the colliding
molecules as rigid rotors.  The coordinate system is defined in a space-fixed
frame, with $\bm{R} \equiv (R,\Omega)$ and $\Omega=(\Phi,\Theta)$ denoting the
intermolecular vector between the centers of mass of the molecules. The coordinates
$\omega_A = (\theta_A,\phi_A)$ and $\omega_B = (\theta_B,\phi_B)$ are the polar
angles of monomers $A$ and $B$, respectively.  We neglect hyperfine structure
and assume that both molecules are in their nuclear-spin stretched states $| I,
M_I=I \rangle$, with $I=I_{\rm{N}}+I_{\rm{H}}=1$ denoting the maximum total
nuclear spin and $M_I$ its laboratory-frame projection.  The scattering
Hamiltonian is then given by
\begin{equation}
\label{eq:H}
\hat{H} = -\frac{\hbar^2}{2\mu R} \frac{\partial^2}{\partial R^2}R +
          \frac{\hat{L}^2}{2\mu R^2} +
          V_S(\bm{R},\omega_A,\omega_B) +
          V_{\rm{magn.dip}}(\bm{R},\hat{S}_A,\hat{S}_B) +
          \hat{H}_A + \hat{H}_B,
\end{equation}
where $\mu$ is the reduced mass of the complex,
$\hat{L}^2$ is the angular momentum operator associated with end-over-end
rotation of the vector $\bm{R}$, $V_S(\bm{R},\omega_A,\omega_B)$ is the potential-energy
surface for total spin $S$, 
$V_{\rm{magn.dip}}(\bm{R},\hat{S}_A,\hat{S}_B)$ is the intermolecular magnetic dipole
interaction, and $\hat{H}_A$ and $\hat{H}_B$ are the Hamiltonians of the two monomers.
The intermolecular magnetic dipole term is given by
\begin{equation}
V_{\rm{magn.dip}}(\bm{R},\hat{S}_A,\hat{S}_B) = - \sqrt{6} g_S^2 \mu_B^2 \frac{\alpha^2}{R^3}
                  \sum_q (-1)^q C_{2,-q}(\Omega) [\hat{S}_A \otimes \hat{S}_B]_q^{(2)}
\end{equation}
where $g_S \approx 2.0023$ is the electron $g$-factor,
$\mu_B$ is the Bohr magneton, $\alpha$ is the fine-structure constant,
and the factor in square brackets is the tensor product of the monomer spin operators 
$\hat{S}_A$ and $\hat{S}_B$.
The monomer operators $\hat{H}_i$ ($i = A, B$) each contain a rotation, spin-rotation, 
intramolecular spin-spin, and Zeeman term,
\begin{equation}
\label{eq:Hi}
\hat{H}_i = B_0 \hat{N}_i^2 + \gamma\hat{N}_i\cdot\hat{S}_i +
            \frac{2}{3}\sqrt{6} \lambda_{\rm{SS}}
            \sum_q (-1)^q C_{2,-q}(\omega_i) [\hat{S}_i \otimes \hat{S}_i]_q^{(2)}
            + g_S \mu_B \bm{B}\cdot\hat{S}_i,
\end{equation}
where $\hat{N}_i$ is the operator for the rotational angular momentum of monomer $i$
and $\bm{B}$ is the magnetic field vector. 
The numerical values for the rotational, spin-rotation, and spin-spin constants of $^{15}$NH 
can be obtained from those of $^{14}$NH \cite{ram:10} using isotope scaling
(see e.g.\ p.\ 239 of Ref.\ \cite{mizushima:75}):
$B_0 = 16.270\,340$ cm$^{-1}$, $\gamma = -0.054\,60$ cm$^{-1}$, and $\lambda_{\rm{SS}} = 0.919\,89$ cm$^{-1}$.

We will assume that all
three spin states of the complex are described by the nonreactive quintet
surface, i.e.\ $V_S(\bm{R},\omega_A,\omega_B) \equiv
V_2(\bm{R},\omega_A,\omega_B)$.  Field-free calculations have shown that this
assumption is reasonable \cite{janssen:10}.  The $S=2$ surface is taken
from Ref.\ \cite{janssen:09}. The potential is expanded in coupled spherical harmonics
$Y_{L,M}$ \cite{green:75},
\begin{align}
\label{eq:V_sf}
V(\bm{R},\omega_A,\omega_B) = {} & \sum_{L_A,L_B,L_{AB}} \upsilon_{L_A,L_B,L_{AB}}(R)
                              A_{L_A,L_B,L_{AB}}(\Omega,\omega_A,\omega_B), \\
\label{eq:A_LLL}
A_{L_A,L_B,L_{AB}}(\Omega,\omega_A,\omega_B) = {} & \sum_{M_A,M_B,M_{AB}}
                            \langle L_A M_A L_B M_B | L_{AB} M_{AB} \rangle \nonumber \\
 & \times  Y_{L_A,M_A}(\omega_A) Y_{L_B,M_B}(\omega_B)  Y^*_{L_{AB},M_{AB}}(\Omega), 
\end{align}
where $\langle L_A M_A L_B M_B | L_{AB} M_{AB} \rangle$ is a Clebsch-Gordan
coefficient and the superscript * denotes complex conjugation.  The subscript
$S=2$ has been omitted for brevity. As noted previously in Ref.\
\cite{janssen:10}, the $\upsilon_{L_A,L_B,L_{AB}}(R)$ expansion
coefficients of Ref.\ \cite{janssen:09} must be multiplied by a factor of
$(-1)^{L_A-L_B} (4\pi)^{-3/2} (2L_{AB}+1) [(2L_A+1)(2L_B+1)]^{1/2}$ to obtain
the potential in the form of Eq.\ (\ref{eq:V_sf}).

We expand the wave function in a space-fixed uncoupled basis introduced in
Ref.\ \cite{krems:04b}. The angular functions are written as products of the
eigenfunctions of $\hat{N}_i^2$, $\hat{N}_{i_z}$,
$\hat{S}_i^2$, $\hat{S}_{i_z}$, $\hat{L}^2$, and $\hat{L}_z$,
\begin{equation}
\label{eq:basis}
| N_A M_{N_A} \rangle | S_A M_{S_A} \rangle
| N_B M_{N_B} \rangle | S_B M_{S_B} \rangle
| L M_L \rangle
\equiv | \gamma_A \gamma_B \rangle | L M_L \rangle,
\end{equation}
with $N_A$ and $N_B$ ranging from $0$ to $N_{\rm{max}}$ and $L = 0, \hdots, L_{\rm{max}}$.
The matrix elements of the scattering Hamiltonian [Eq.\ (\ref{eq:H})] in the
basis of Eq.\ (\ref{eq:basis}) have all been given by Krems and Dalgarno
\cite{krems:04b}. Note that the factor of $g_S^2 \mu_B^2 \approx 1.0023$ $e^2\hbar^2/m_e^2$ 
has been neglected in their expression for the magnetic dipole interaction.
Furthermore,
it should be taken into account that our potential expansion [Eq.\
(\ref{eq:V_sf})] differs by a factor of $(-1)^{L_A-L_B} (4\pi)^{3/2}
(2L_{AB}+1)^{1/2}$ from the expansion used in Ref.\ \cite{krems:04b}.

Since the monomers are identical, we can exploit the permutation
symmetry of the system to minimize the computational cost. Following Ref.\
\cite{tscherbul:09c}, we employ a normalized, symmetrized basis of the form
\begin{equation}
\label{eq:basis_sa}
| \phi^{\eta}_{\gamma_A \gamma_B L M_L} \rangle = 
\frac{1}{[2(1+\delta_{\gamma_A \gamma_B})]^{1/2}} 
\big[
| \gamma_A \gamma_B \rangle | L M_L \rangle 
+ \eta (-1)^L
| \gamma_B \gamma_A \rangle | L M_L \rangle 
\big],
\end{equation}
with $\eta=\pm1$ defining the symmetry of the wave function with respect to
molecular interchange. 
It is sufficient to restrict the basis to a well-ordered
set of molecular states, i.e.\ $\gamma_A \geq \gamma_B$.  Finally, we note that
the parity, $\epsilon = (-1)^{N_A + N_B + L}$, and the total angular momentum
projection quantum number, $\mathcal{M} = M_{N_A} + M_{S_A} + M_{N_B} + M_{S_B}
+ M_L$, are also conserved during collision.  Thus, the scattering calculations
may be performed for a single value of $\mathcal{M}$, $\eta$, and $\epsilon$.
Here we consider only collisions between bosonic $^{15}$NH molecules
in their spin-stretched and rotational ground state ($N_A=N_B=0$), for which we
have $\eta=+1$ and $\epsilon=+1$.  Note that the first excited rotational state
has an energy of $\approx32$ cm$^{-1}$ (46 K), and is therefore inaccessible at
the energies considered in this work.

We solve the coupled equations using the hybrid log-derivative method of
Alexander and Manolopoulos \cite{alexander:87}, which uses a fixed-step-size
log-derivative propagator in the short range and a variable-step-size Airy
propagator at long range. Matching to asymptotic boundary conditions yields the
scattering $S$-matrices, from which the cross sections can be readily obtained.
We note that, due to the intramolecular spin-rotation and spin-spin couplings,
the basis functions of Eq.\ (\ref{eq:basis_sa}) are not exact eigenfunctions of
the asymptotic Hamiltonian $\hat{H}_A + \hat{H}_B$, while the $S$-matrices must be constructed in terms
of these eigenfunctions. As detailed in Refs.\ \cite{krems:04b} and
\cite{tscherbul:09c}, an additional basis transformation of the log-derivative matrix is therefore
required before matching to the asymptotic boundary conditions.
We will denote the symmetrized channel eigenfunctions as 
$| \phi^{\eta}_{\bar{\gamma}_A \bar{\gamma}_B L M_L} \rangle$, with
$| \bar{\gamma}_A \bar{\gamma}_B \rangle \equiv
| (\bar{N}_A,S_A)J_A, M_{J_A} \rangle | (\bar{N}_B,S_B)J_B, M_{J_B} \rangle $ defining
the molecular eigenstates. Here $\bar{N}_i$ is used instead of $N_i$, because $N_i$ is
strictly not an exact quantum number due to the intramolecular spin-spin coupling.
However, the intramolecular coupling
is relatively weak and $N_i, M_{N_i}$, and $M_{S_i}$ may be treated as
approximately good quantum numbers. Specifically, for the $^{15}$NH rotational ground
state, the magnetically trapped $| J_i=1, M_{J_i}=1 \rangle$ component 
contains 99.992\% of $| N_i=0, M_{N_i}=0, S_i=1, M_{S_i}=1 \rangle$ for all fields considered in this work.

The cross sections at total energy $E$ are calculated using the expression
\cite{tscherbul:09c}
\begin{equation}
\label{eq:xsec}
\sigma^{\eta}_{\bar{\gamma}_A \bar{\gamma}_B \rightarrow \bar{\gamma}'_A \bar{\gamma}'_B}(E) =
\frac{\pi (1+\delta_{\bar{\gamma}_A \bar{\gamma}_B})}{k^2_{\bar{\gamma}_A \bar{\gamma}_B}}
\sum_{L,M_L} \sum_{L', M'_L}
\left| T^{\eta}_{\bar{\gamma}_A \bar{\gamma}_B L M_L;\bar{\gamma}'_A \bar{\gamma}'_B L' M'_L}(E) \right|^2,
\end{equation}
where $k_{\bar{\gamma}_A \bar{\gamma}_B}$ is the wavenumber for the incident channel,
$k^2_{\bar{\gamma}_A \bar{\gamma}_B} = 2\mu(E - \epsilon_{\bar{\gamma}_A} - \epsilon_{\bar{\gamma}_B})/\hbar^2$,
$\epsilon_{\bar{\gamma}_A} + \epsilon_{\bar{\gamma}_B}$ is the corresponding channel energy, and
the $T$-matrix elements are defined as $T^{\eta}_{\bar{\gamma}_A \bar{\gamma}_B L M_L;
\bar{\gamma}'_A \bar{\gamma}'_B L' M'_L} = \delta_{\bar{\gamma}_A \bar{\gamma}'_A}
\delta_{\bar{\gamma}_B \bar{\gamma}'_B} \delta_{L L'} \delta_{M_L M_L'} -
S^{\eta}_{\bar{\gamma}_A \bar{\gamma}_B L M_L;\bar{\gamma}'_A \bar{\gamma}'_B L' M'_L}$. 

\subsection{Computational details}
We performed the scattering calculations for different magnetic fields using
the MOLSCAT package \cite{molscat:94,gonzalez:07}, extended to handle the
basis set of Eq.\ (\ref{eq:basis_sa}). The log-derivative propagation was
carried out on a radial grid ranging from $4.5$ to $15$ $a_0$ in steps of
$0.02$ $a_0$.  The Airy propagation ranged from $15$ to $50\,000$ $a_0$. We
included basis functions up to $N_{\rm{max}} = 2$ and $L_{\rm{max}} = 6$, yielding a maximum
number of 1038 channels for a single calculation ($\mathcal{M}=0$). The expansion
of the quintet potential-energy surface was truncated at $L_A = L_B = 6$.
As mentioned in Section \ref{subsec:theory}, the chemically reactive
singlet and triplet potentials were not included in the calculations.

\section{Results and discussion}
\label{sec:results}
\subsection{Adiabatic potential-energy curves}
\label{subsec:avcrossings}
\begin{figure}
\centering
\includegraphics[width=15cm]{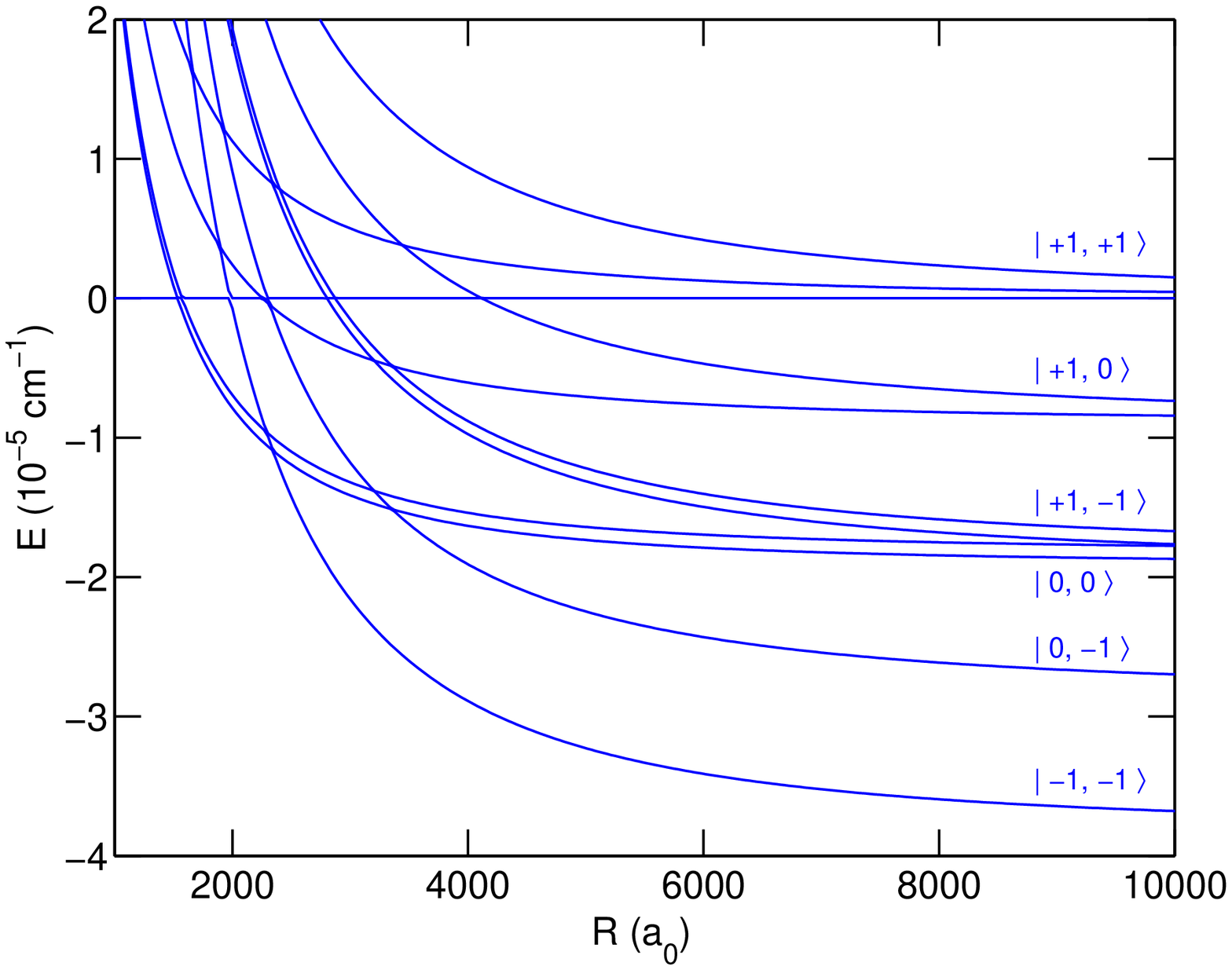} 
\caption
  {\label{fig:adiab}
Lowest adiabatic potential curves for $^{15}$NH -- $^{15}$NH, calculated
for $L_{\rm{max}} = 4$ and $\mathcal{M} = 2$ at a magnetic field of 0.1 G. 
The molecular eigenstates are labeled by $| M_{J_A}, M_{J_B} \rangle$ and refer
to the well-ordered states $| (N_A=0)J_A=1, M_{J_A} \rangle | (N_B=0)J_B=1, M_{J_B} \rangle$.
}
\end{figure}

Before discussing the cross sections, we first consider the adiabatic potential
curves. These are obtained by diagonalizing the interaction matrix at fixed $R$ over a grid
of $R$-values and subsequently connecting the corresponding eigenvalues.
Asymptotically, these adiabatic curves correlate to the molecular eigenstates
$\bar{\gamma}_A$ and $\bar{\gamma}_B$. Figure \ref{fig:adiab} shows the long
range of the lowest adiabats for $\mathcal{M} = 2$ and $L_{\rm{max}} = 4$ at a
magnetic field of 0.1 G.
We present only the curves with exchange symmetry
$\eta=+1$ and parity $\epsilon=+1$, for which $s$-wave scattering in the incident
spin-stretched channel is allowed. The asymptotic energy splittings
between the different molecular states originate from the Zeeman term $g_S \mu_B B
(M_{S_A} + M_{S_B})$. For $\mathcal{M}=2$ and $L_{\rm{max}}=4$, the magnetically trapped state with
$|M_{J_A}=M_{J_B}=1\rangle$ has partial-wave contributions of $L=0$, 2, and 4, as can be seen in Fig.\ \ref{fig:adiab}.
The adiabats correlating with $|M_{J_A}=1,M_{J_B}=0\rangle$, $|M_{J_A}=1,M_{J_B}=-1\rangle$, and
$|M_{J_A}=0,M_{J_B}=0\rangle$ have $L=2$ and 4 centrifugal barriers, and the adiabats for
$|M_{J_A}=0,M_{J_B}=-1\rangle$ and $|M_{J_A}=-1,M_{J_B}=-1\rangle$ contain only
the $L=4$ partial wave. For the $| M_{J_A}=1, M_{J_B}=-1 \rangle$ and $|
M_{J_A}=0, M_{J_B}=0 \rangle$ states, which have identical Zeeman shifts, the 
degeneracy is further lifted by the intramolecular spin-spin coupling.

It can be seen that several curve crossings
occur in the region between $R \approx 1500$ and 4500 $a_0$. If we neglect
the small energy shifts due to the intramolecular coupling, the
crossing points $R_c$ are simply given by
\begin{equation}
g_S \mu_B B \Delta M_S = \frac{\hbar^2 \left[L_f(L_f+1) - L_i(L_i+1)\right] }{2\mu R_c^2},
\end{equation}
where $L_i$ and $L_f$ denote the values of $L$ for the adiabats correlating to the incoming and outgoing
channels, respectively, and $\Delta M_S = M_{S_A}^{(i)}+M_{S_B}^{(i)} -
M_{S_A}^{(f)}-M_{S_B}^{(f)}$. The corresponding energies at which the crossings
occur are given by $E_c = \hbar^2 L_i(L_i+1)/(2\mu R_c^2)$,
defined relative to the threshold of the incident channel.
We find that several of these crossings are narrowly
avoided due to the presence of the intermolecular magnetic dipole interaction.
Inspection of Eq.\ (30) of Ref.\ \cite{krems:04b} or Eq.\ (A2) of
Ref.\ \cite{hutson:08} shows that e.g.\ the $s$-wave incident channel with
$| M_{S_A} = 1, M_{S_B} = 1, L = 0 \rangle$ is directly coupled with
$| M_{S_A} = 1, M_{S_B} = 0, L = 2 \rangle$ and
$| M_{S_A} = 0, M_{S_B} = 0, L = 2 \rangle$ via the magnetic dipole term,
giving rise to the corresponding avoided crossings. We note, however, that
$V_{\rm{magn.dip}}$ contains a second-rank tensor in $\Omega$ and can therefore
directly couple channels only if $|L_i-L_f| \leq 2$. The spin operators
$\hat{S}_A$ and $\hat{S}_B$ contained in $V_{\rm{magn.dip}}$ are first-rank tensors and, consequently,
$M_{S_A}$ and $M_{S_B}$ may each differ at most by 1.
Thus, not all crossings are avoided. To take proper account of these avoided crossings
and account correctly for the magnetic dipole coupling in the ultracold regime, it is
essential that the radial grid used in the calculations extends beyond the outermost $R_c$ value.
The influence of the avoided crossings on the collision cross sections will be
discussed in detail in the following section.


\subsection{Cross sections}
\begin{figure}
\centering
\includegraphics[width=15cm]{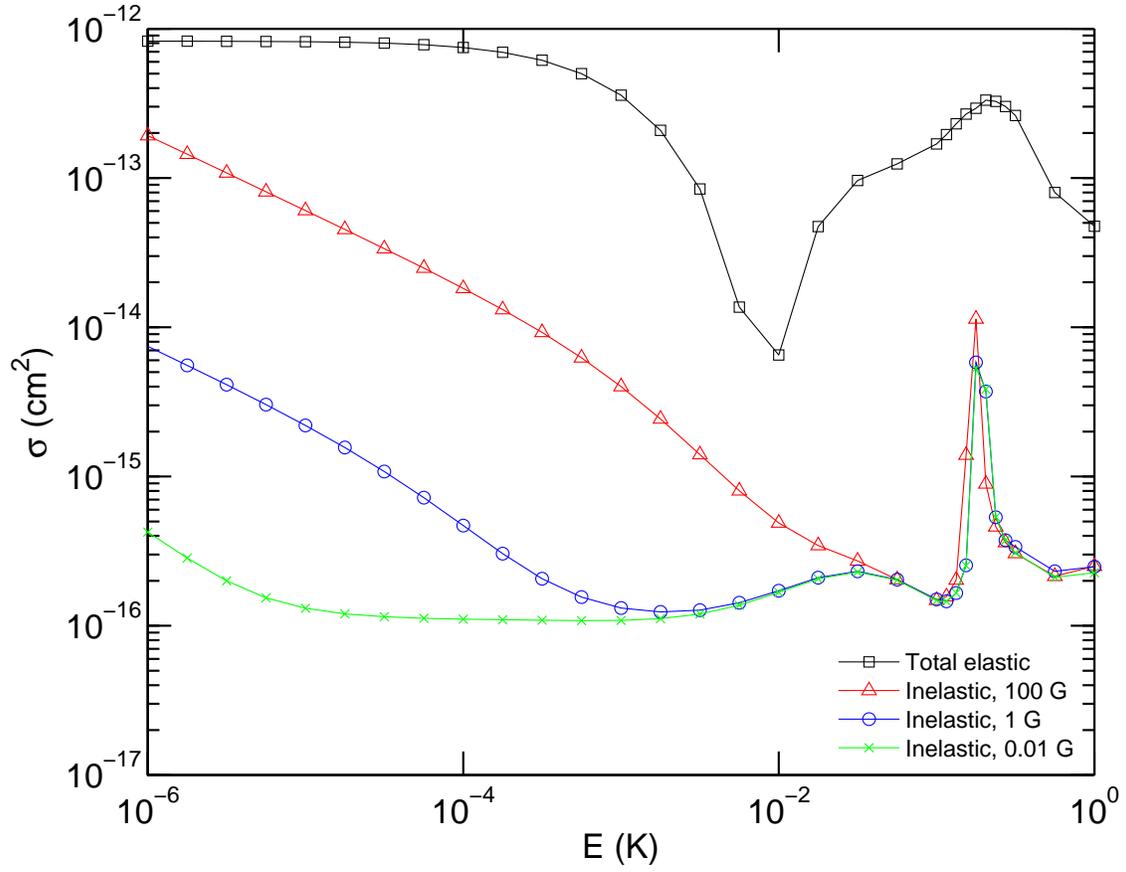} 
\caption
  {\label{fig:xsecE}
Elastic and inelastic $M_{J}$-changing cross sections for magnetically trapped
$^{15}$NH as a function of collision energy for various magnetic fields. The elastic
cross sections are the same for all three magnetic field strengths.
}
\end{figure}

\begin{figure}
\centering
\includegraphics[width=15cm]{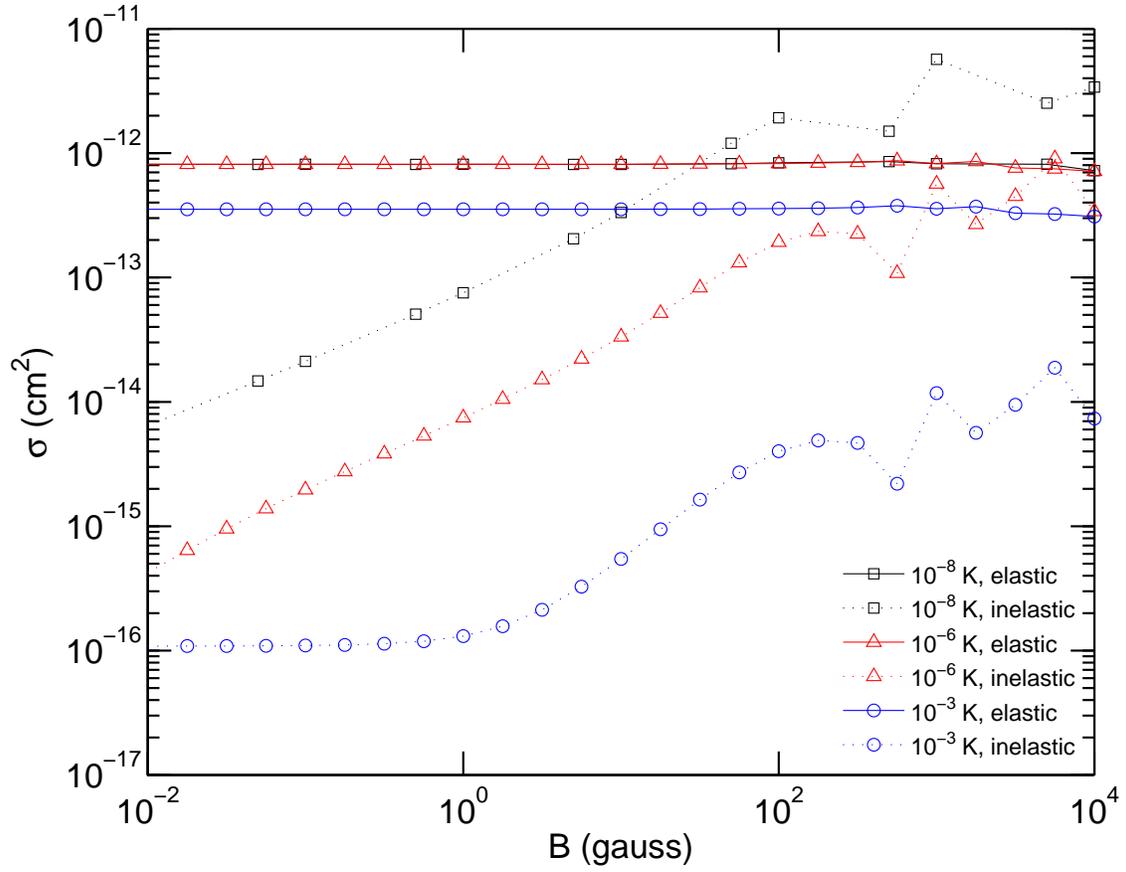} 
\caption
  {\label{fig:xsecB}
Elastic and inelastic $M_{J}$-changing cross sections for magnetically trapped
$^{15}$NH as a function of magnetic field for various collision energies.
}
\end{figure}

Figure \ref{fig:xsecE} shows the total elastic and $M_J$-changing inelastic cross sections
for two magnetically trapped $^{15}$NH molecules ($M_{J_A} = M_{J_B} = 1$) as a
function of collision energy. The cross sections as a function of magnetic field are
given in Fig.\ \ref{fig:xsecB}. We find that the elastic cross sections are significantly
larger than the inelastic ones over a wide range of energies and
magnetic fields, suggesting that evaporative cooling of $^{15}$NH is likely to
be successful. It can also be seen that, in the ultracold regime, the total
inelastic cross section decreases dramatically if the magnetic field strength
is reduced. Thus, once the cooling process has started in the millikelvin regime
at relatively high magnetic field, and continues towards lower energies as the magnetic
trap depth is decreased, the ratio between elastic and $M_J$-changing cross sections
will remain very favorable for evaporative cooling to take place.

%

\begin{figure}
\centering
\includegraphics[width=15cm]{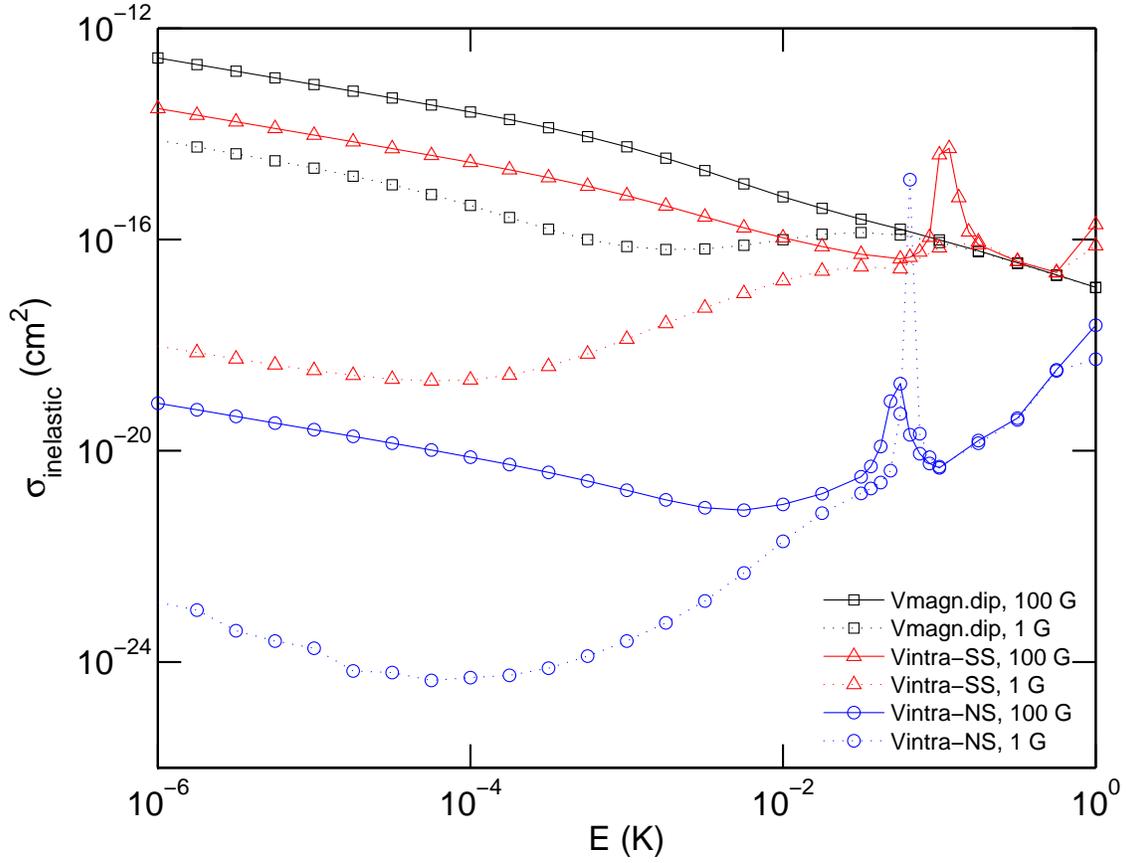} 
\caption
  {\label{fig:xsecE_inelas}
Inelastic $M_{J}$-changing cross sections for $\mathcal{M}=2$ as a
function of collision energy at 1 G and 100 G.
The different curves correspond to calculations with either
the intermolecular magnetic dipolar coupling (``$V_{\rm{magn.dip}}$"), or the intramolecular
spin-spin coupling (``$V_{\rm{intra-SS}}$"), or the intramolecular spin-rotation coupling (``$V_{\rm{intra-NS}}$")
included in $\hat{H}$.
}
\end{figure}

\begin{figure}
\centering
\includegraphics[width=15cm]{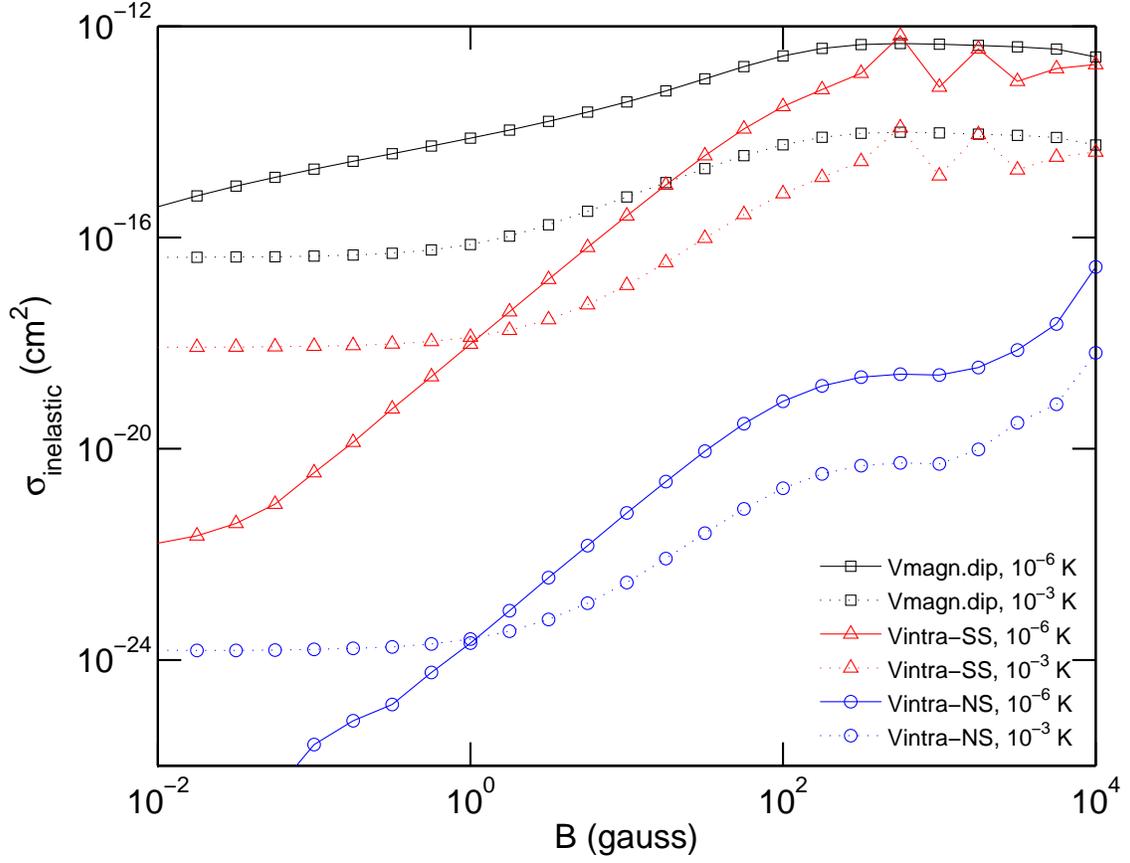} 
\caption
  {\label{fig:xsecB_inelas}
Inelastic $M_{J}$-changing cross sections for $\mathcal{M}=2$ as a
function of magnetic field at 10$^{-6}$ K and 10$^{-3}$ K.
The different curves are obtained from scattering calculations with either
the intermolecular magnetic dipolar coupling (``$V_{\rm{magn.dip}}$"), or the intramolecular
spin-spin coupling (``$V_{\rm{intra-SS}}$"), or the intramolecular spin-rotation coupling (``$V_{\rm{intra-NS}}$")
switched on.
}
\end{figure}

In order to identify the main trap-loss mechanism, we have also performed
scattering calculations where two of the three spin-dependent coupling terms
are set to zero. 
The results for $\mathcal{M}=2$ are shown in Figs.\ \ref{fig:xsecE_inelas} and \ref{fig:xsecB_inelas}.
We have verified that the $\mathcal{M}=2$ cross sections, which allow for 
$s$-wave collisions in the incident channel, are dominant at the energies
and fields considered in this work.
It can be seen
that the intermolecular spin-spin interaction ($V_{\rm{magn.dip}}$) gives the
largest contribution to the inelastic cross section over a broad range of
energies and field strengths, most notably in the ultracold regime.
This is also the case for cold and ultracold N + NH collisions \cite{zuchowski:10,hummon:10}.
At higher collision energies and fields,  
however, the \textit{intramolecular} coupling terms become increasingly
important. In particular the intramolecular spin-spin term causes significant
trap loss above $E \approx 10^{-2}$ K and $B \approx 100$ G.
Spin-rotation coupling, which vanishes for pure $N_A=N_B=0$ states,
has only a very small effect on the total cross section. This is consistent with
previous work on the He--NH($^3\Sigma^-$) system \cite{krems:04b, krems:03a, cybulski:05}.


%

The importance of the \textit{intermolecular} magnetic dipole interaction is
most easily understood by considering the adiabatic potential curves. As
discussed in Section \ref{subsec:avcrossings}, the avoided crossings between
the $s$-wave incoming channel $| M_{J_A}=1, M_{J_B}=1, L=0 \rangle$ and the $|
M_{S_A} = 1, M_{S_B} = 0, L = 2 \rangle$ and $| M_{S_A} = 0, M_{S_B} = 0, L = 2
\rangle$ outgoing channels
all occur at very long range for small to moderate field strengths.
Consequently, the spin-flip induced by $V_{\rm{magn.dip}}$ can take place without
having to overcome the $d$-wave 
barrier in the outgoing channels, and hence
this process dominates the inelastic cross section in the ultracold regime.
We also emphasize that, if the avoided-crossing points $R_c$ fall outside the
scattering propagation region, i.e.\ if the radial grid is chosen too small,
the inelastic cross sections are similar to the case where the intermolecular
magnetic dipole term is switched off completely. This confirms that the spin-flip
due to $V_{\rm{magn.dip}}$ indeed takes place at long range, or more specifically,
at $R \approx R_c$.

Kajita \cite{kajita:06} has shown that, in the Born approximation, the
inelastic cross section $\sigma_{i\rightarrow f}$ for $^{15}$NH ($L_i=0,
L_f=2$) caused by the magnetic dipole interaction should be proportional to
$B^{1/2}$ and $E^{-1/2}$ if the kinetic energy release is relatively large
($k_i \ll k_f$). This is consistent with our results obtained from full quantum
scattering calculations in the ultracold regime.
As the collision energy increases, however, the assumption of $k_i \ll k_f$ breaks down
and the cross sections deviate from the $B^{1/2}$ behaviour. We find numerically that this
is the case for energies above $\sim 10^{-6}$ K at nearly all the field strengths considered
in this work (see Fig.\ \ref{fig:xsecB}).
In a separate publication, we will give a general
analytical expression for the inelastic cross section due to $V_{\rm{magn.dip}}$
based on the (distorted-wave) Born approximation, and show that the numerical and analytical results 
are in excellent agreement over a wide range of energies and fields.

At collision energies above about 10 mK or high magnetic fields, there is sufficient
energy to overcome the $d$-wave centrifugal barrier in the outgoing channels
and, as a consequence, short-range effects become important. In particular the
\textit{intramolecular} spin-spin coupling term, which requires strong anisotropy of the
potential in order to induce Zeeman relaxation \cite{krems:04b,krems:03a}, enhances
the inelastic cross section significantly at energies above $\sim$ 10 mK. 
For the intramolecular contributions, we find that the inelastic cross section behaves as $B^{5/2}$ at
moderate field strengths and flattens off to a constant value at very small
$B$. Its kinetic energy dependence is proportional to $E^{-1/2}$ in the
ultracold regime and, for small magnetic fields, also shows a
region of $E^2$ behaviour. This result is consistent with the work of Volpi
and Bohn, who applied the distorted-wave Born approximation to calculate
inelastic spin-changing collisions induced at short range \cite{volpi:02}:
\begin{equation}
\sigma_{i \rightarrow f} \propto E^{L_i - 1/2} \left( E + g_S \mu_B B \Delta M_S \right) ^{L_f + 1/2}.
\end{equation}
Thus, for an $s$-wave incoming channel ($L_i = 0$) and $d$-wave outgoing
channel ($L_f = 2$), there is a region of $E^2$ dependence when the Zeeman
shift for the outgoing channel is small compared to the collision energy. At
very small fields, the magnetic-field dependence is negligible and the cross
section becomes constant as a function of $B$.

\begin{figure}
\centering
\includegraphics[width=15cm]{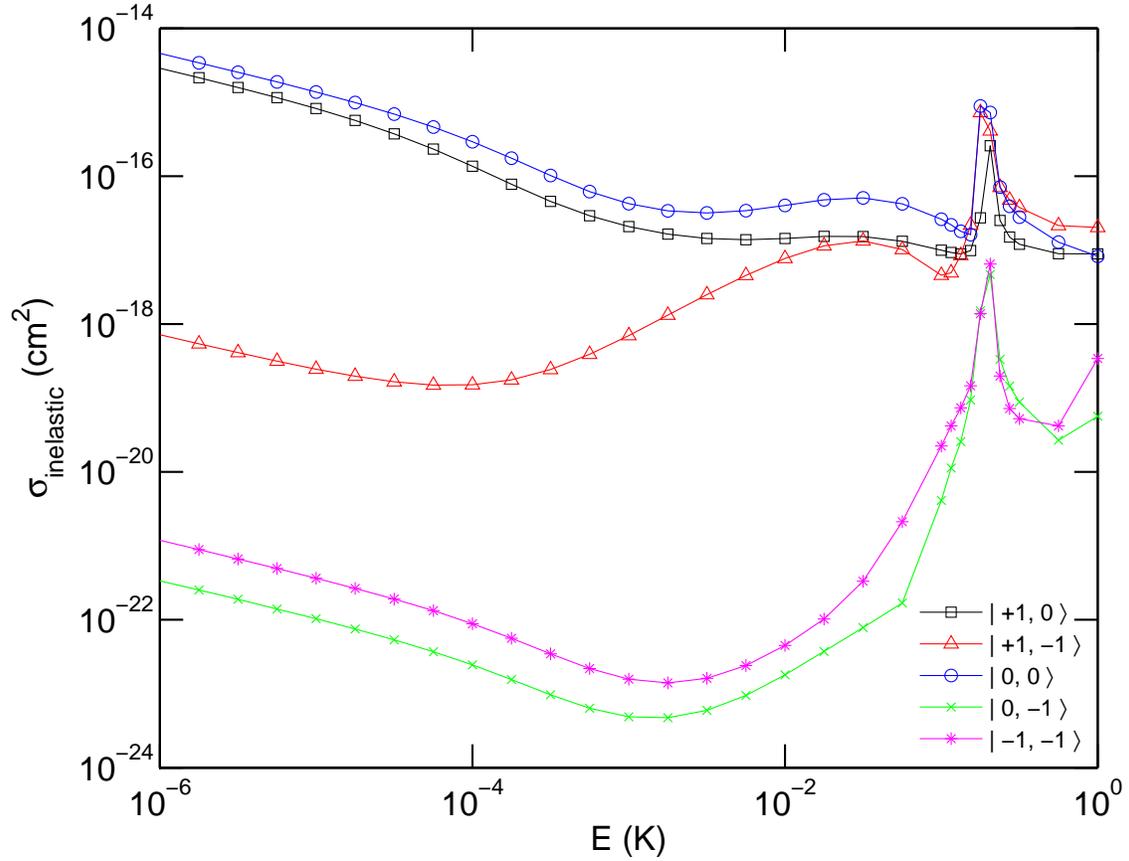} 
\caption
  {\label{fig:xsec_ss}
State-to-state inelastic cross sections ($\mathcal{M} = 2$) for magnetically trapped $^{15}$NH
as a function of energy at 1 G. 
The final states are labeled by $| M_{J_A}, M_{J_B} \rangle$.
}
\end{figure}

We also point out that, in contrast to the \textit{intermolecular} spin-spin
coupling term, the \textit{intramolecular} spin-spin interaction contains second-rank
tensors in $\hat{S}_A$ and $\hat{S_B}$ and therefore directly
couples channels where $M_{S_A}$ and $M_{S_B}$ each differ by 0, 1, \textit{or}
2. Thus, transitions from $M_{S_i} = 1$ to $M_{S_i} = -1$
also become allowed in first order.  This is illustrated in Fig.\ \ref{fig:xsec_ss}, where the
state-to-state inelastic cross sections for $\mathcal{M}=2$ are plotted as a
function of energy. In the ultracold region, which is dominated by
$V_{\rm{magn.dip}}$, only the $| M_{S_A} = 1, M_{S_B}=1\rangle$ $\rightarrow$ $|
M_{S_A} = 1, M_{S_B}=0 \rangle$ and $| M_{S_A} = 1, M_{S_B}=1 \rangle
\rightarrow | M_{S_A} = 0, M_{S_B}=0 \rangle$ transitions contribute to the
inelastic cross section. At higher energies, where the intramolecular spin-spin
term plays a significant role, transitions from \mbox{$| M_{S_A} = 1, M_{S_B}=1, L=0 \rangle$} to the
\mbox{$| M_{S_A} = 1, M_{S_A} = -1, L=2 \rangle$}, \mbox{$| M_{S_A} = 0, M_{S_A} = -1, L =4
\rangle$}, and \mbox{$| M_{S_A} = -1, M_{S_A} = -1, L=4 \rangle$} channels
become increasingly important. Note that the latter two have $g$-wave barriers
in the exit channel, and hence they are strongly suppressed in the low-energy
regime.


In summary, we have established that the dominant trap-loss mechanism for NH 
in the ultracold regime is the \textit{intermolecular} spin-spin coupling term, which
induces Zeeman relaxation at \textit{long range}. 
When the kinetic energy in the outgoing channel becomes large, the spin-change is also
caused by the interplay of the potential anisotropy and the \textit{intramolecular} spin-spin interaction,
which acts at \textit{short range}.

\subsection{Dependence on potential and basis-set size}

\begin{figure}
\centering
\includegraphics[width=15cm]{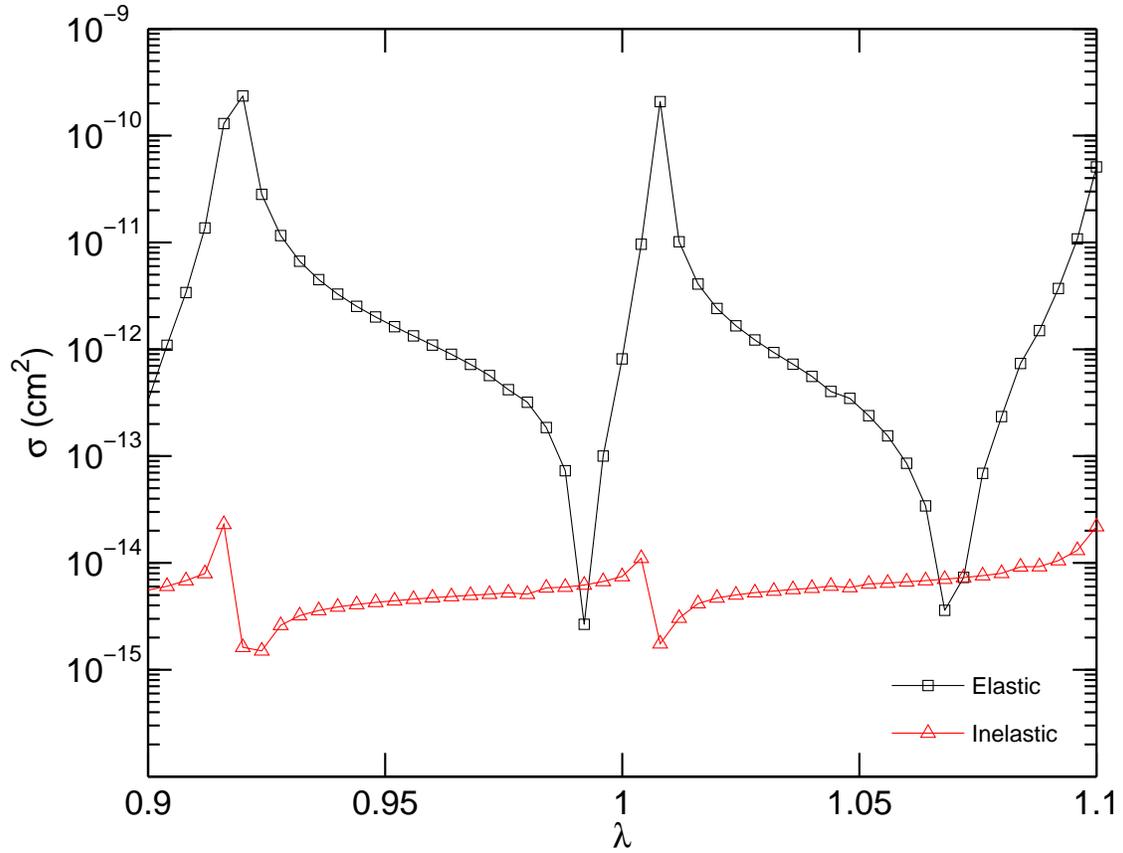} 
\caption
  {\label{fig:xsecmu}
Elastic and inelastic $M_{J}$-changing cross sections ($\mathcal{M}=2$) for 
magnetically trapped $^{15}$NH as a function of the scaling factor $\lambda$,
calculated at a collision energy of $10^{-6}$ K and a magnetic field strength of 1 G.
}
\end{figure}

As a final part of our discussion, we consider the sensitivity to the potential and
the dependence on the size of the channel basis set. It is well known that low-energy
collisions are highly sensitive to the exact form of the interaction potential,
so that we must carefully take into account the uncertainty in the potential-energy
surface. We estimate that the quintet potential used in the scattering
calculations is accurate to within a few percent.  Following Refs.\
\cite{zuchowski:10} and \cite{janssen:10}, we have studied the potential
dependence indirectly by evaluating the sensitivity to the reduced mass. The
reduced mass was scaled by a factor of $\lambda$ (\mbox{$0.9 \leq \lambda \leq 1.1$}),
which is essentially equivalent to scaling of the entire potential by $\lambda$
\cite{zuchowski:10}. Figure \ref{fig:xsecmu} shows the $\mathcal{M}=2$ cross
sections for different scaling factors at a collision energy of 10$^{-6}$ K and
a field strength of 1 G.
We find that the cross sections exhibit resonance structures at certain $\lambda$ values,
indicating a high sensitivity to the potential. However, the ratio between elastic and inelastic
cross sections is much less sensitive to the reduced mass, and the prospects for evaporative cooling
remain very positive over almost the entire range of $\lambda$. The contributions from the different
spin-changing mechanisms, as described in the previous section, are also very similar for different
reduced masses. Our qualitative results and conclusions are thus
reasonably independent of the precise form of the potential.

\begin{figure}
\centering
\includegraphics[width=15cm]{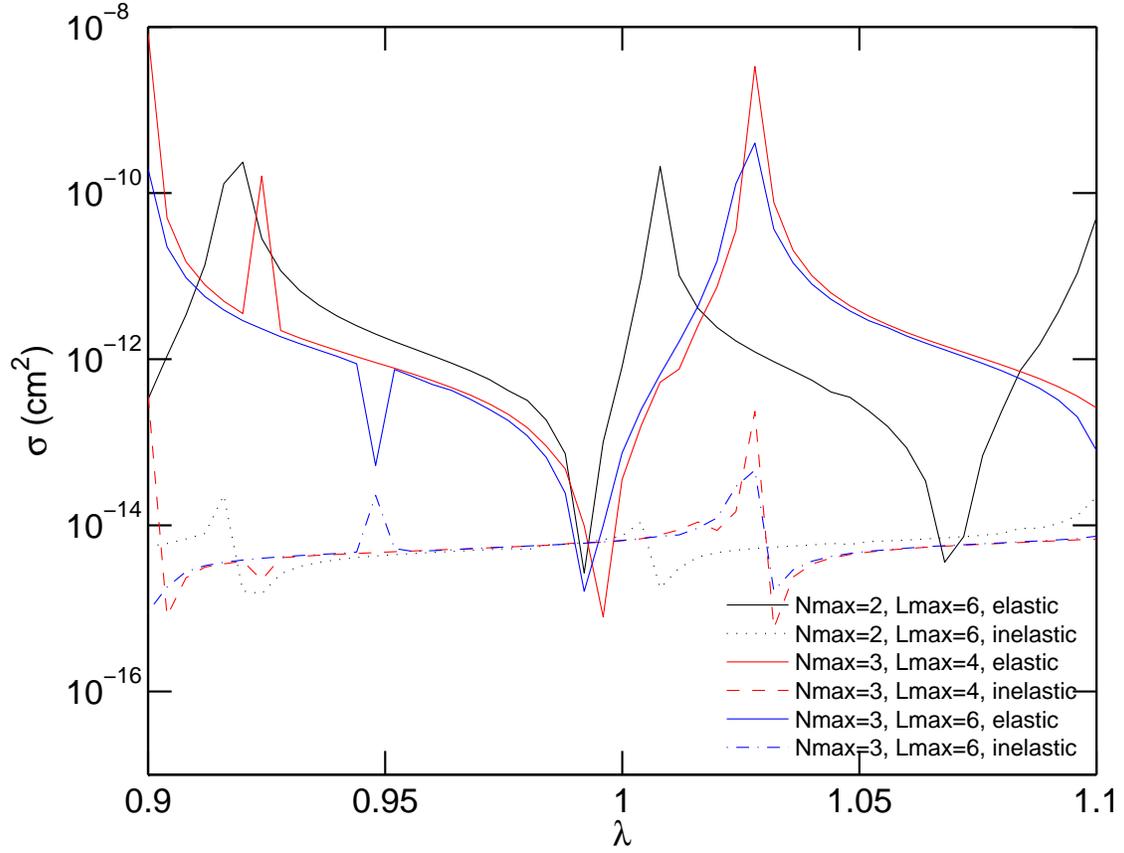} 
\caption
  {\label{fig:xsecmu_basis}
Elastic and inelastic $M_{J}$-changing cross sections ($\mathcal{M}=2$)
as a function of the scaling factor $\lambda$, calculated for different basis sets
at a collision energy of $10^{-6}$ K and a magnetic field strength of 1 G.
}
\end{figure}

We use the same scaling approach to investigate the dependence on basis-set size.
Figure \ref{fig:xsecmu_basis} shows the $\mathcal{M}=2$ cross sections as a function of $\lambda$
for different values of $N_{\rm{max}}$ and $L_{\rm{max}}$ at $E = 10^{-6}$ K and $B = 1$ G.
The total number of channels in these calculations ranged from 901 for $N_{\rm{max}}=2, L_{\rm{max}}=6$
up to 2621 for $N_{\rm{max}}=3, L_{\rm{max}}=6$.
It can be seen that 
the positions of the resonances shift when the basis size is increased, but
the general pattern remains essentially the same. This is consistent with other work on NH
\cite{janssen:10,zuchowski:10,wallis:10}.
We thus conclude that, within the uncertainty limits
of $\lambda$,
our qualitative results are not very sensitive to the size of the angular basis set.

Figure \ref{fig:xsecmu_basis} also demonstrates that the cross sections are not
yet converged with respect to $N_{\rm{max}}$ and $L_{\rm{max}}$.  In fact, field-free
NH--NH calculations suggest that the basis set should extend to at least
$N_{\rm{max}}=6$ and $L_{\rm{max}}=7$ to achieve full convergence \cite{janssen:10},
which would amount to a maximum of 
25\,598 channels for $\mathcal{M}=0$ in the
present decoupled basis set.  Such calculations are highly infeasible with the
currently available computer power.  However, taking into account the
uncertainty in the potential and the pronounced resonance structure, even a
fully converged basis set would not give really reliable numerical values.
Nonetheless, it must be emphasized that unconverged basis sets can give
qualitatively reliable results.
As already discussed in Refs.\ \cite{janssen:10} and \cite{wallis:10}, full basis-set
convergence is therefore not strictly necessary within the uncertainty limits
of the potential.


\section{Conclusions}

We have carried out full quantum scattering calculations to study cold and
ultracold $^{15}$NH -- $^{15}$NH collisions in magnetic fields. The elastic and
spin-changing cross sections for magnetically trapped NH are found to be very
favorable for efficient evaporative cooling. We have identified the
\textit{intermolecular} spin-spin coupling term as the main trap-loss mechanism
at low energies and small magnetic fields, while the \textit{intramolecular}
spin-spin term becomes increasingly important at higher energies and fields.
The effect of spin-rotation coupling is almost negligible.

Finally, we have demonstrated that the numerical values of the cross sections
are very sensitive to the details of the potential, but the
\textit{qualitative} results and conclusions are almost independent of the
exact form of the surface. The size of the angular basis set, which is almost
impossible to converge for systems such as NH--NH, does not significantly alter
the results within the uncertainty limits of the potential. This inherent
uncertainty in the calculated cross sections, however, clearly highlights the need for
reliable experimental data.

\begin{acknowledgments}
We gratefully acknowledge EPSRC for funding the collaborative project CoPoMol
under the ESF EUROCORES programme EuroQUAM. LMCJ and GCG thank the Council for
Chemical Sciences of the Netherlands Organization for Scientific Research
(CW-NWO) for financial support.
\end{acknowledgments}

\bibliography{vanderwaals}

\clearpage

\clearpage

\clearpage

\clearpage

\clearpage

\clearpage

\clearpage

\clearpage

\end{document}